\documentclass[a4paper,10pt]{article}

\usepackage[dvips]{graphicx}
\usepackage{epsfig,amsmath,amssymb,verbatim,mathrsfs,array,layout,textcomp,amssymb,latexsym}

\newcommand{\beq}{\begin{eqnarray}}
\newcommand{\eeq}{\end{eqnarray}}

\def\beqa{\begin{eqnarray}}
\def\eeqa{\end{eqnarray}}
\newcommand{\no}{\nonumber}
\newcommand{\bv}{\left(\begin{array}{c}}
\newcommand{\ev}{\end{array}\right)}
\newcommand{\bmtwo}{\left(\begin{array}{cc}}
\newcommand{\bmthree}{\left(\begin{array}{ccc}}
\newcommand{\emn}{\end{array}\right)}
\newcommand{\bmtwoc}{\left\{\begin{array}{cc}}
\newcommand{\bmthreec}{\left\{\begin{array}{ccc}}
\newcommand{\emnc}{\end{array}\right\}}
\newcommand{\ba}{\begin{array}}
\newcommand{\ea}{\end{array}}

\newcommand{\lag}{\mathcal{L}}

\def\lsim{\mathrel{\rlap{\lower4pt\hbox{\hskip1pt$\sim$}}
     \raise1pt\hbox{$<$}}}         
\def\gsim{\mathrel{\rlap{\lower4pt\hbox{\hskip1pt$\sim$}}
     \raise1pt\hbox{$>$}}}         

\begin{document}

\begin{titlepage}

\vskip1.5cm
\begin{center}
  {\Large \bf Scalar-mediated $t\bar t$ forward-backward asymmetry\\ }
\end{center}
\vskip0.2cm

\begin{center}
{\bf Kfir Blum, Yonit Hochberg and Yosef Nir}

\end{center}
\vskip 8pt

\begin{center}
{\it Department of Particle Physics and Astrophysics\\
Weizmann Institute of Science,\\ Rehovot 76100, Israel} \vspace*{0.3cm}

{\tt  kfir.blum,yonit.hochberg,yosef.nir@weizmann.ac.il}
\end{center}

\vglue 0.3truecm

\begin{abstract}
  \vskip 3pt \noindent A large forward-backward asymmetry in $t\bar t$
  production, for large invariant mass of the $t\bar t$ system, has been
  recently observed by the CDF collaboration. Among the scalar
  mediated mechanisms that can explain such a large asymmetry, only
  the $t$-channel exchange of a color-singlet weak-doublet scalar is
  consistent with both differential and integrated $t\bar t$ cross section measurements. Constraints from flavor changing processes dictate a very specific structure for the Yukawa couplings of such a new scalar. No sizable deviation in the differential or integrated $t\bar t$ production cross section is expected at the LHC.
\end{abstract}

\end{titlepage}

\section{Introduction}
\label{sec:intro}
The CDF collaboration has recently observed a large forward-backward
$t\bar t$ production asymmetry for large invariant mass of the $t\bar
t$ system~\cite{Aaltonen:2011kc}:
\beq\label{eq:atthexp}
A^{t\bar t}_h\equiv A^{t\bar t}(M_{t\bar t}\geq450\ {\rm
  GeV})=+0.475\pm0.114\,,
\eeq
to be compared with the Standard Model (SM)
prediction~\cite{Almeida:2008ug, Bowen:2005ap, Antunano:2007da},
$\left(A^{t\bar t}_h\right)_{SM}=+0.09\pm0.01$. Eq.~(\ref{eq:atthexp})
updates (and is consistent with) previous CDF and D0 measurements of
the inclusive asymmetry~\cite{:2007qb,Aaltonen:2008hc}. Such a large
effect is suggestive of an interference effect between a tree level
exchange of a new boson with an electroweak-scale mass and the SM
gluon-mediated amplitude. The intermediate boson could be either a
vector-boson or a scalar. In this work, we focus on the latter
possibility
\cite{Grinstein:2011yv,Patel:2011eh,Ligeti:2011vt,AguilarSaavedra:2011vw,Gresham:2011pa,Shu:2011au,AguilarSaavedra:2011zy,Nelson:2011us,Zhu:2011ww,Babu:2011yw,Cui:2011xy,AguilarSaavedra:2011ug,Vecchi:2011ab}.
(For earlier work on scalar mediated mechanisms that give a large
inclusive forward-backward asymmetry, see Refs.
\cite{Shu:2009xf,Arhrib:2009hu,Dorsner:2009mq,Jung:2009pi,Cao:2009uz,Cao:2010zb}.)

Eight scalar representations can interfere with the SM in $t\bar t$
production at the Tevatron:
\beq\label{eq:scarep}
(\bar 6,1)_{-\frac{4}{3}},\;\;(\bar 6,1)_{-\frac{1}{3}},\;\;(\bar 6,3)_{-\frac{1}{3}},\;\;
(3,1)_{-\frac{4}{3}},\;\;(3,1)_{-\frac{1}{3}},\;\;(3,3)_{-\frac{1}{3}},\;\;
(8,2)_{-\frac{1}{2}},\;\;(1,2)_{-\frac{1}{2}}.
\eeq
In Section \ref{sec:top} we argue that only the color-singlet weak
doublet $(1,2)_{-\frac{1}{2}}$ can enhance $A^{t\bar t}_h$ without
being inconsistent with existing measurements of the total $t\bar t$
production cross section or invariant mass distribution.

Guided by the top-related collider measurements, we focus our
attention on the weak doublet. Given that $\mathcal{O}(1)$ coupling is
required to either $u\bar t$ or $d\bar t$, the weak doublet cannot be
the ordinary Higgs doublet, nor can it be incorporated into a two
Higgs doublet model with ``natural flavor conservation''. It is thus
clear that the scalar sector of the theory has a special flavor
structure. In this work, we emphasize the interplay between the features
required to explain the top-related measurements and the flavor
constraints.

The plan of the sections investigating the color-singlet weak-doublet
scalar is as follows. In Section~\ref{sec:doublet} we introduce our
notation and formalism for the investigation of the extra weak
doublet. In Section~\ref{sec:flavor} we show how flavor-related
observables allow for only a very specific flavor structure of the
scalar doublet. The implications for electroweak precision
measurements are studied in Section~\ref{sec:ewpm}, and for additional
top-related observables in Section~\ref{sec:topsss}. We summarize our
results in Section~\ref{sec:sum}.

\section{Top-related constraints}
\label{sec:top}
When new physics is invoked to account for the large value of
$A^{t\bar t}_h$, one has to make sure that such new physics does not
violate the constraints from other top-related measurements.
Specifically, we consider the following measurements:

(i) The forward-backward $t\bar t$ production asymmetry for small
invariant mass of the $t\bar t$ system~\cite{Aaltonen:2011kc}:
\beq\label{eq:attlexp}
A^{t\bar t}_l\equiv A^{t\bar t}(M_{t\bar t}\leq450\ {\rm
  GeV})=-0.116\pm0.153\,,
\eeq
to be compared with the SM prediction~\cite{Almeida:2008ug},
$A^{t\bar t}_l=+0.040 \pm0.006$.

(ii) The $t\bar t$ differential cross section~\cite{Aaltonen:2009iz}.
We represent this constraint by considering a particular $M_{t\bar t}$
bin:
\beqa\label{eq:higsig}
\sigma_h&\equiv& \sigma^{t\bar t}(700\ {\rm GeV}<M_{t\bar
  t}<800\ {\rm GeV})=80\pm37\ {\rm fb}\,,
\eeq
to be compared with the SM
prediction~\cite{Almeida:2008ug,Ahrens:2010zv}, $\sigma_h=80\pm8$ fb.

(iii) The $t\bar t$ total production cross section measured
at CDF~\cite{cdfinc},
\beqa\label{eq:incsig}
\sigma_i&\equiv& \sigma^{t\bar t}_{\rm tot}=7.50\pm0.48\ {\rm pb}\,,
\eeq
consistent with D0 measurements~\cite{d0inc}. Some controversy exists
regarding the theoretical SM prediction. Ref.~\cite{Kidonakis:2010dk}
obtains $\sigma_i=7.2\pm0.4$ pb, consistent with previous
results~\cite{incth}, but in some disagreement with
Ref.~\cite{Ahrens:2010zv} which obtains $\sigma_i=6.5\pm0.3$ pb. In
what follows, we conservatively allow a $^{+30\%}_{-10\%}$ uncertainty on the
$t\bar t$ total production cross section.

For each of the eight representations of Eq. (\ref{eq:scarep}), only
few parameters are relevant to the calculation of the $t\bar t$
production cross section and forward-backward asymmetry. Typically,
these parameters include the mass $M$, a coupling $|\lambda|^2$, and
the width $\Gamma$. In some cases, there is more than one coupling of
relevance, forming a slightly more involved parameter space. In
practice, in all cases we find that varying the width between
$\Gamma=(0.01-0.5)M$ does not affect the results significantly. The
most essential features, such as the sign of the leading interference
terms with the SM diagrams and the possibility of forward-peaking
kinematical features, are dictated purely by the choice of
representation. This situation enables us to compute the top-related
observables for each of the representations, and check for consistency
with collider data, in a model-independent way.

The details of our calculation for each representation are given in
App.~\ref{app:ope}. We work at LO, using the MSTW2008 LO PDF
set~\cite{Martin:2009iq} and adopting renormalization scale and
factorization scale $\mu_R=\mu_F=\sqrt{\tilde s}$, where $\tilde s$ is
the partonic center of mass energy. Where not stated otherwise, we use
$\alpha_S(m_Z)=0.139$. We estimate the uncertainties in our
calculation by varying the renormalization and factorization scales
within the range $(0.5\sqrt{\tilde s}-2\sqrt{\tilde s})$, by comparing
the results to results obtained using the CTEQ5M PDF
set~\cite{Lai:1999wy}, and by varying the value of $\alpha_S(m_Z)$ in
the range $(0.117-0.139)$. The largest uncertainties we find arise
from varying $\alpha_S$, leading to an uncertainty on $A^{t\bar t}$ which can be as large $10\%-30\%$, but is typically smaller. We checked that these uncertainties do not affect our conclusions. In order to
minimize the impact of NLO corrections to the new physics (NP)
contributions, we normalize the new physics contribution to the SM
one~\cite{Blum:2011up}. We assume that the $K$-factors are universal,
so that the NP/SM ratios at LO and NLO are the same.

We (conservatively) consider a parameter region as ruled out if any of
the following conditions applies:
\beqa\label{eq:ruleout}
A_{h,{\rm NP}}^{t\bar t}&<&0.2,\no\\
A_{l,{\rm NP}}^{t\bar t}&>&0.2,\no\\
N_{h}\equiv \left|\sigma_{h}^{\rm NP}\right|/\sigma_{h}^{\rm SM}&>&1,\no\\
N_{i}\equiv\sigma_{i}^{\rm
    NP}/\sigma_{i}^{\rm SM}&>&+0.3\;\;{\rm or}\;<-0.1.
\eeqa

Our main result is that, except from the color-singlet weak-doublet
$(1,2)_{-{1}/{2}}$, all of the other scalar representations are ruled
out from explaining the large forward-backward asymmetry
Eq.~(\ref{eq:atthexp}). The reason is that all of the colored
representations enhance the $t\bar t$ production cross section at low
and/or high invariant mass, in conflict with experimental data. Thus,
if the parameters relevant for these representations are tuned to pass
the first criterion in Eq.~(\ref{eq:ruleout}), they inevitably fail
on at least one of the last two criteria.

For the color sextet and triplet, this tension was mentioned in Refs.
\cite{Grinstein:2011yv} and \cite{Ligeti:2011vt}, respectively, but
was not taken to imply that the models are excluded. It was also
pointed out in Ref. \cite{Gresham:2011pa}.

In Table~\ref{tab:t} we present, for each representation, the largest
value of $A_{h,{\rm NP}}^{t\bar t}$ which passes each of the remaining
criteria of Eq.~(\ref{eq:ruleout}). A failure on any of these criteria
is manifest through a value of $A_{h,{\rm NP}}^{t\bar t}<0.2$. A
failure of a certain representation to account for the
forward-backward asymmetry is manifest through at least one entry in
the three columns that has $A_{h,{\rm NP}}^{t\bar t}<0.2$.
\begin{table}[h!]
  \caption{The maximum value of $A_{h,{\rm NP}}^{t\bar t}$ for a given
    top-related constraint.}
\label{tab:t}
\begin{center}
\begin{tabular}{c|cccc} \hline\hline
  \rule{0pt}{1.2em}%
 Irrep &  max$\left\{A_{h,{\rm NP}}^{t\bar t}\,\Big|\,N_h<1\right\}$ &
 max$\left\{A_{h,{\rm NP}}^{t\bar t}\,\Big|\,-0.1<N_i<+0.3\right\}$ &
 max$\left\{A_{h,{\rm NP}}^{t\bar t}\,\Big|\,A_{l,{\rm NP}}^{t\bar t}<0.2\right\}$ \cr \hline
$(1,2)_{-\frac{1}{2}}$ (u) & 0.27 & 0.25& 0.29\cr\hline
$(1,2)_{-\frac{1}{2}}$ (d)& 0.17 & 0.09& 0.24\cr
$(8,2)_{-\frac{1}{2}}$ (u)& 0.08 & 0.02& 0.23\cr
$(8,2)_{-\frac{1}{2}}$ (d)& 0.12 & 0.03& 0.23\cr
$(\bar 6,1)_{-\frac{1}{3}}$ & 0.17 & 0.19& 0.44\cr
$(\bar 6,3)_{-\frac{1}{3}}$ & 0.14 & 0.18& 0.49\cr
$(\bar 6,1)_{-\frac{4}{3}}$ & 0.14 & 0.18& 0.50\cr
$(3,1)_{-\frac{1}{3}}$      & 0.17 & 0.27& 0.44\cr
$(3,3)_{-\frac{1}{3}}$      & 0.11 & 0.42& 0.55\cr
$(3,1)_{-\frac{4}{3}}$      & 0.10 & 0.41& 0.55\cr
\hline\hline
\end{tabular}
\end{center}
\end{table}

\begin{figure}[h!]\begin{center}
\includegraphics[width=8cm]{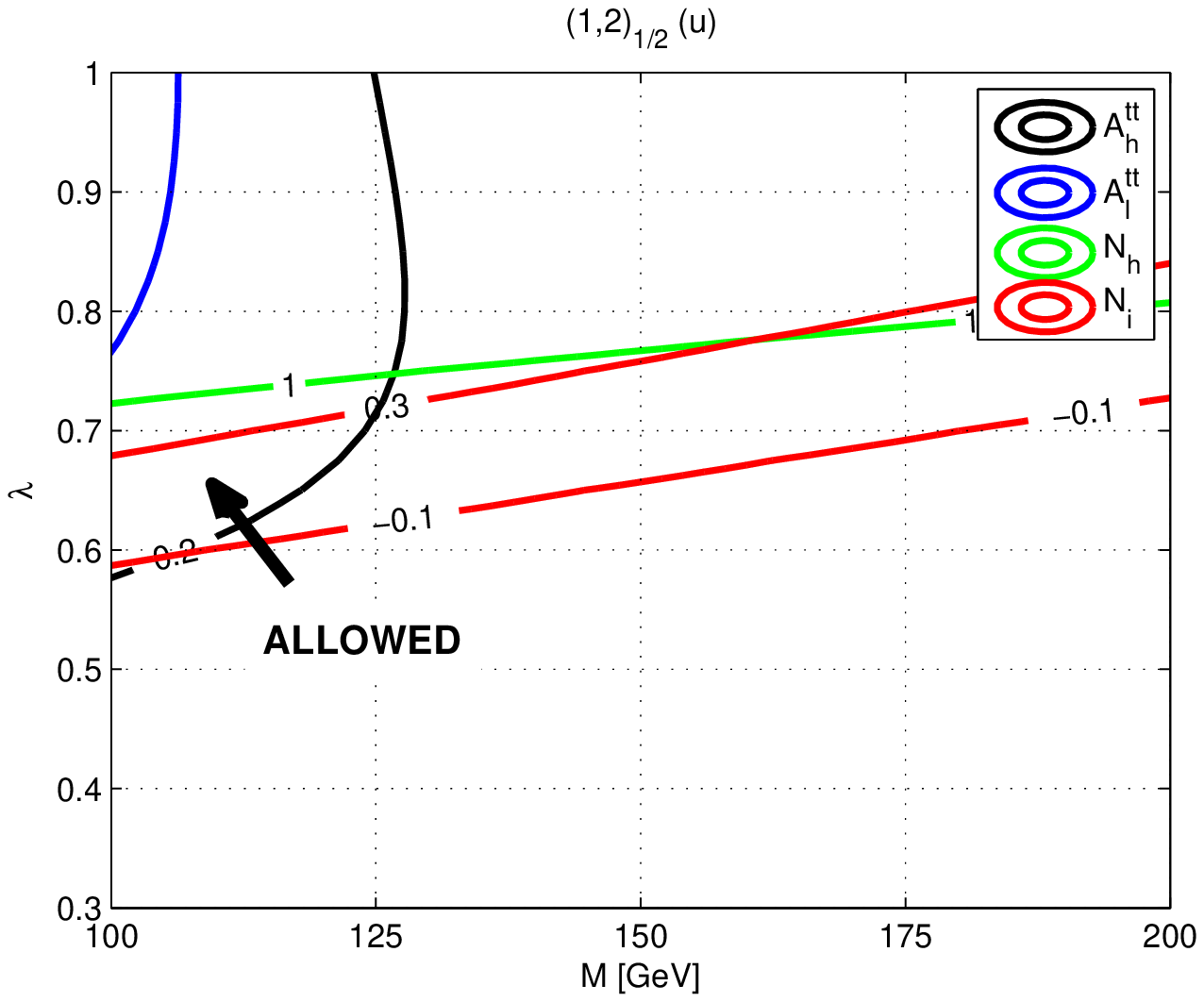}\hfill
\includegraphics[width=8cm]{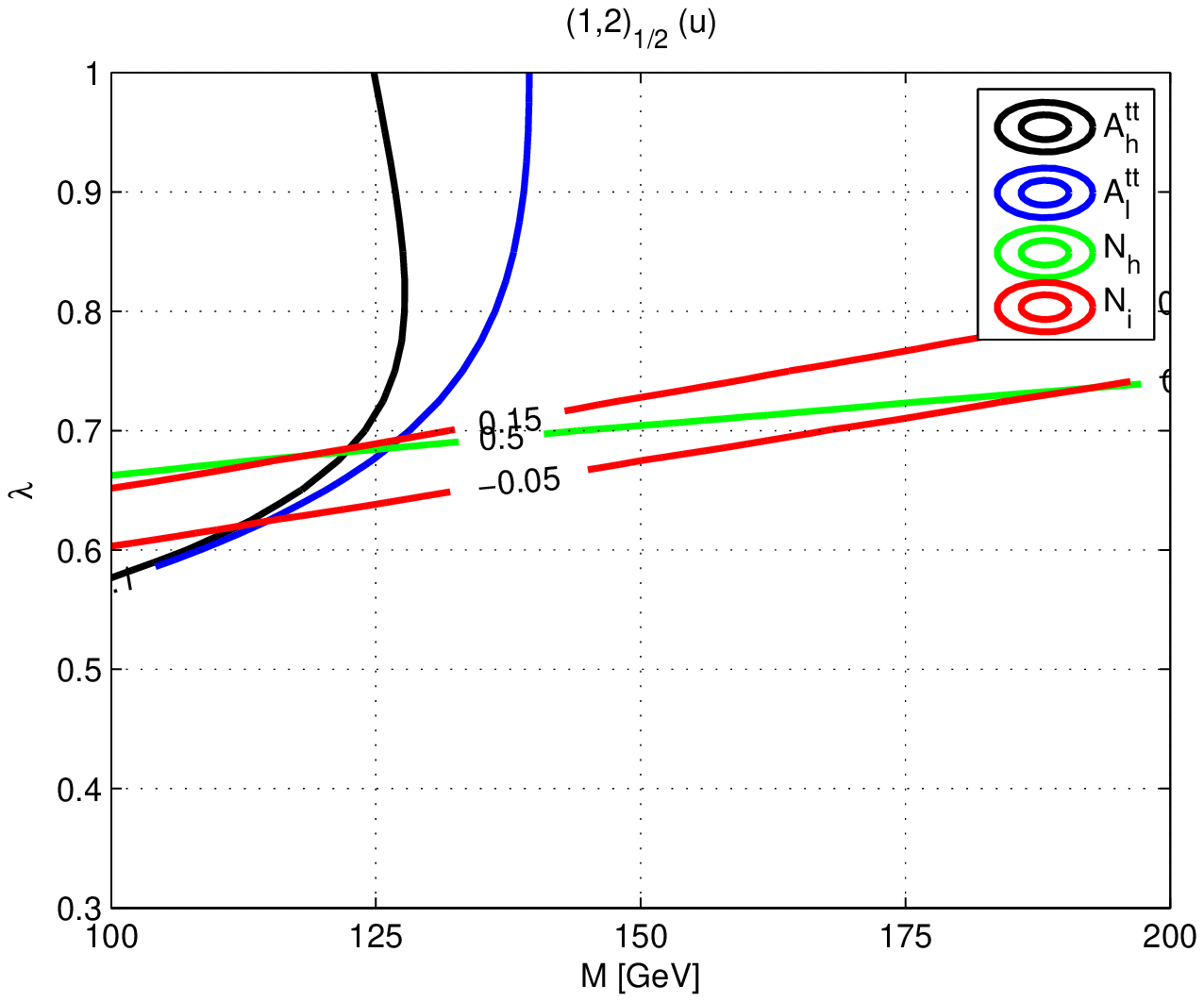}
\caption{The $(M,|\lambda|)$ parameter space for the
  $(1,2)_{-{1}/{2}}$ scalar, coupling the top to the up quark. Black
  (blue) curves correspond to $A_{h}^{t\bar t}(A_{l}^{t\bar t})$.
  Green (red) curves correspond to $N_h(N_i)$. Left panel: Parameter
  space allowed by Eq.~(\ref{eq:ruleout}). Here, the area below the
  black curve, the area above the blue curve, the area above the green
  curve and the area above the upper and inside the lower red curves
  are ruled out by $>2\sigma$ by the corresponding observable. Right
  panel: Same as on the left, but limiting the deviations on
  $A_l^{t\bar t},\,N_h$ and $N_i$ to $\lsim1\sigma$. No parameter region satisfies $A_h^{t\bar t}>0.2$ and $A_l^{t\bar t}<0.1$ while simultaneously satisfying cross section constraints at one sigma.}
\label{fig:12a}
\end{center}
\end{figure}

The small $(1,2)_{-{1}/{2}}$ parameter region which survives all of
the conditions of Eq.~(\ref{eq:ruleout}) is shown in the left panel of
Fig.~\ref{fig:12a}. Here, the scalar couples the top to up quarks (see
Section \ref{sec:doublet} for details on the scalar couplings). The
allowed region, where large $A_{h}^{t\bar t}$ can be obtained without
violating cross section constraints, has $M<130$ GeV and
$\mathcal{O}(1)$ coupling. (The fact that a color-singlet
  weak-doublet scalar can explain the CDF value of $A_h^{t\bar t}$ was first pointed out in Ref.~\cite{AguilarSaavedra:2011vw}. They use methods of effective field theory, and thus do not derive bounds on the mass and couplings of the scalar.) Our results are consistent with the results of Refs.
\cite{Nelson:2011us,Babu:2011yw,Cui:2011xy}. The allowed parameter
region extends to low scalar mass, without conflict with $t\bar t$
cross section measurements, due to interference with the SM $s-$channel
gluon exchange diagram. The interference reduces the forward-backward
symmetric and enhances the asymmetric production cross section.
Considering LEPII searches we limit the discussion to scalar mass
larger than 100 GeV. For $M$ below $m_t$, top decay to the weak-doublet can effect the parameter space depicted in Fig.~\ref{fig:12a}; See discussion in Section~\ref{sec:tdec}.

In the right panel of Fig.~\ref{fig:12a} we plot $A_h^{t\bar t}$ together with contours corresponding roughly to the $1\sigma$ allowed ranges for the cross sections and low bin asymmetry. We see that the low bin asymmetry places the tightest constraint on the doublet model. No parameter region is found with $A_h^{t\bar t}>0.2$ and $A_l^{t\bar t}<0.1$.

At the Tevatron, if the coupling involves right-handed up quarks
(u-type), then $t\bar t$ production could feed off the $u\bar u$
and/or the $d\bar d$ luminosity. On the other hand, if the coupling
involves right-handed down quarks (d-type), then $t\bar t$ production
must feed off the $d\bar d$ luminosity. Since the $d\bar d$ luminosity
makes up only $\lsim15\%$ of the $u\bar u$ luminosity at high
invariant mass, producing $A_h^{t\bar t}$ via the d-type coupling
requires a larger coupling for a given value of the scalar mass. Thus,
the d-type coupling is more tightly constrained by the $t\bar t$ cross
section measurements and, as can be seen in Table~\ref{tab:t}, it
fails the $N_i$ and $N_h$ criteria of Eq.~(\ref{eq:ruleout}).

Despite this tension of d-type coupling with the $t\bar t$ cross
section, in what follows we keep the d-type coupling in the
discussion. We do this mainly to cover the possibility that both u-
and d-type couplings occur in conjunction, leading to potential
combined effects that could in principle broaden the allowed parameter
range. Eventually, we will show that flavor constraints preclude this
possibility.

It was pointed out in Refs.~\cite{Gresham:2011pa,Jung:2011zv} that models in which $t\bar t$ production proceeds via $t-$channel exchange can have sizable corrections to the parton level $t\bar t$ cross section, as deduced by CDF. While such acceptance issues are not relevant for the color sextet and triplet representations, the cross section constraints quoted here for the color singlet and octet could be somewhat over restrictive.\footnote{We thank Jure Zupan and Alex Kagan for discussions on this point.} For the color singlet, these effects might extend the allowed parameter region depicted in Fig.~\ref{fig:12a} to include slightly larger values of $M$ and $\lambda$. For the color octet, we have checked that even relaxing the total cross section constraint by a factor of three, would not allow to explain $A_h^{t\bar t}$ with $M>100$~GeV.

Finally, we comment that the same interference mechanism which helps the
color-singlet weak-doublet scalar evade the Tevatron constraints on
the $t\bar t$ production cross section is at work also at the LHC. Thus,
no sizable distortion of the differential or inclusive $t\bar t$
production cross section is expected at the LHC. For instance, taking $\lambda=0.7$ and $M=120$~GeV, at the border of the allowed parameter space defined in Fig.~\ref{fig:12a}, produces only a $\sim 30\%$ deviation in the differential $t\bar t$ cross section at $M_{t\bar t}=1.5$~TeV at the LHC.

\section{The weak doublet}
\label{sec:doublet}
We consider a weak doublet
\beq
\Phi\sim(1,2)_{-{1}/{2}}=\bv\phi^0\\ \phi^-\ev.
\eeq
This color-singlet can couple left-handed quarks to right-handed
quarks in either the up or the down sector.
%

\subsection{Coupling to right-handed up quarks}
The relevant Lagrangian terms in the quark mass basis are given by
\beq\label{eq:defx}
\lag_u=-V(\Phi)+\left[
  2\phi^0u_{Li}^\dag X_{ij}u_{Rj}+2\phi^-d_{Li}^\dag (V^\dag
  X)_{ij}u_{Rj}+{\rm h.c.}\right],
\eeq
where $X$ is a complex $3\times3$ matrix in flavor space
and $V=V_{\rm CKM}$.

To account for the forward-backward asymmetry in $t\bar t$ production,
$\Phi$ needs to have an ${\cal O}(1)$ coupling to $u\bar t$. There are
two such possibilities: $X_{13}={\cal O}(1)$ and/or $X_{31}={\cal O}(1)$.
Indeed, from Fig.~\ref{fig:12a} we learn that $A_{h}^{t\bar t}>0.2$
implies
\beq\label{eq:afbcstrt}
|\lambda|>0.6,\;\;\;M<130\,{\rm GeV},
\eeq
where $\lambda$ refers to either $X_{13}$ or $X_{31}$. We could expect
that a smaller $\lambda$ is allowed in case that both $X_{13}$ and
$X_{31}$ exist with similar magnitude. We will see, however, that
flavor constraints imply $|X_{13}|\ll1$, and so in practice this case
cannot occur.

Since a left-handed quark doublet is involved in the Lagrangian
(\ref{eq:defx}), there is often an interesting interplay between
flavor constraints from the up and the down sector (see, for example,
\cite{Blum:2009sk}). In what follows, in order to explore this
interplay, we consider four different structures for $X$:
\beq\label{eq:x31}
X=\lambda\left(\ba{ccc}0&0&0\\0&0&0\\1 &0&0\ea\right),\ \ \
V^\dagger X=\lambda\left(\ba{ccc}V_{td}^*&0&0\\V_{ts}^*&0&0\\
  V_{tb}^*&0&0\ea\right)
\;\;\;\;&{\rm case\, I},\\
X=\lambda\left(\ba{ccc}V_{ub}&0&0\\V_{cb}&0&0\\V_{tb}&0&0\ea\right),\ \ \
V^\dagger X=\lambda\left(\ba{ccc}0&0&0\\0&0&0\\1&0&0\ea\right)
\;\;\;\;&{\rm case\, II},\label{eq:vx31}\\
X=\lambda\left(\ba{ccc}0&0&1\\0&0&0\\ 0&0&0\ea\right),\ \ \
V^\dagger X=\lambda\left(\ba{ccc}0&0&V_{ud}^*\\0&0&V_{us}^*\\
  0&0&V_{ub}^*\ea\right)
\;\;\;\;&{\rm case\, III},\label{eq:x13}\\
X=\lambda\left(\ba{ccc}0&0&V_{ud}\\0&0&V_{cd}\\0&0&V_{td}\ea\right),\ \ \
V^\dagger X=\lambda\left(\ba{ccc}0&0&1\\0&0&0\\ 0&0&0\ea\right)
\;\;\;\;&{\rm case\, IV}.\label{eq:vx13}
\eeq
When there is a single entry in $X$, $\phi^0$ couples to a single pair
of up-type (mass eigenstate) quarks, but $\phi^-$ couples an up-type
quark to all three down-type quarks. In contrast, when there is a
single entry in $V^\dagger X$, $\phi^-$ couples to a single up-down
pair, while $\phi^0$ couples a single right-handed up-type quark to
all three left-handed up-type quarks. Consequently, a single entry in
$X$ is subject to flavor constraints from the down sector, while a
single entry in $V^\dagger X$ is subject to flavor constraints from
the up sector.

In general, if both case I and case II are excluded, then so will be
all intermediate cases, where both $X_{31}$ and $(V^\dagger X)_{31}$
are ${\cal O}(1)$. Similarly, in general, if both case III and case IV
are excluded, then so will be all intermediate cases, where both
$X_{13}$ and $(V^\dagger X)_{13}$ are ${\cal O}(1)$. Exceptions to
these statements might arise in cases that there are cancellations
between various contributions to flavor changing processes. Such
cancellations indeed occur in MFV models~\cite{KaZu}.

\subsection{Coupling to right-handed down quarks}
The relevant Lagrangian terms in the quark mass basis are given by
\beq\label{eq:deftx}
\lag_d=-V(\Phi)+\left[
  2\phi^+u_{Li}^\dag \widetilde X_{ij}d_{Rj}-2\phi^{0*}d_{Li}^\dag (V^\dag
  \widetilde X)_{ij}d_{Rj}+{\rm h.c.}\right],
\eeq
where $\widetilde X$ is a complex $3\times3$ matrix in flavor space
and $V=V_{\rm CKM}$.

Here there are only two different structures for $\widetilde X$ to consider:
\beq\label{eq:tilx31}
\widetilde X=\widetilde\lambda\left(\ba{ccc}0&0&0\\0&0&0\\1 &0&0\ea\right),\ \ \
V^\dagger \widetilde X=\widetilde\lambda\left(\ba{ccc}V_{td}^*&0&0\\V_{ts}^*&0&0\\
  V_{tb}^*&0&0\ea\right)
\;\;\;\;&{\rm case\, }{\rm V},\\
\widetilde X=\widetilde\lambda\left(\ba{ccc}V_{ub}&0&0\\V_{cb}&0&0\\V_{tb}&0&0\ea\right),\ \ \
V^\dagger \widetilde X=\widetilde\lambda\left(\ba{ccc}0&0&0\\0&0&0\\1&0&0\ea\right)
\;\;\;\;&{\rm case\,}{\rm VI},\label{eq:tilvx31}
\eeq
analogous to Eqs.~\eqref{eq:x31} (case I) and \eqref{eq:vx31} (case
II) above. Structures analogous to the cases III and IV in
Eqs.~\eqref{eq:x13} and \eqref{eq:vx13} are irrelevant here as they do
not couple the top quark to the light quarks.

To generate a sizable forward-backward asymmetry in $t\bar t$ production,
the coupling between $\Phi$ and $d\bar t$ should be of order one,
$\widetilde\lambda={\cal O}(1)$. As seen in Table~\ref{tab:t}, however, if the Lagrangian contains only the coupling in Eq.~(\ref{eq:deftx}) then a large forward-backward asymmetry cannot be attained without violating cross section constraints.

\section{Flavor constraints}
\label{sec:flavor}
In this section, we denote the masses of both the neutral and charged
components of the scalar doublet by $M$. As will be seen in
Section~\ref{ssec:t}, the electroweak $T$-parameter constrains
the splitting within the doublet such that the use of single scale $M$
is valid for our discussion of flavor constraints.

\subsection{Meson mixing}\label{ssec:mix}
Scalar exchange can contribute to neutral meson mixing at tree level
and via box diagrams. However, in all the cases that we consider, the
constraints from tree level exchange are irrelevant. We thus focus on
the loop contributions.  These include box diagrams with two scalars
and in some cases box diagrams with one scalar and one gauge or
goldstone boson. In all of the cases relevant for our scenario, the
mixed scalar-gauge diagrams, if exist, are parametrically suppressed
by either a small quark mass insertion or by CKM factors compared to the
diagrams with two internal scalars.

The strongest constraints on our scenarios arise from the $K$ and $D$
systems. Using the analysis of Ref.~\cite{Isidori:2010kg}, we obtain
the following constraints on (the absolute value of) our model
parameters:
\begin{itemize}
\item $D^0-\overline{D}{}^0$ mixing:
\beq\label{eq:d}
\frac{1}{32\pi^2}\left(\frac{M}{\rm TeV}\right)^{-2}
\sum_i\mathcal{F}\left(\frac{m_{u_i}^2}{M^2}\right)X^2_{1i}X^{*2}_{2i}
<7\times10^{-7}.
\eeq
\item $K^0-\overline{K}{}^0$ mixing:
\beq\label{eq:k}
  \frac{1}{32\pi^2}\left(\frac{M}{\rm TeV}\right)^{-2}
  \sum_i\mathcal{F}\left(\frac{m_{u_i}^2}{M^2}\right)
  (V^\dag X)^2_{1i}(V^\dag X)^{*2}_{2i}
  <10^{-6}.
\eeq
\end{itemize}
The loop function $\mathcal{F}$ is given by
\beq\label{eq:Fkkdd}
\mathcal{F}(r)&=&\frac{1-r^2+2r\ln r}{(1-r)^3}.
\eeq
The function $\mathcal{F}$ obeys $\mathcal{F}(1)=\frac{1}{3}$,
$\mathcal{F}(0)=1$.  Constraints on the couplings of Eq.
(\ref{eq:deftx}) are obtained by replacing $m_{u_i}\to m_{d_i}$ and
$X\to \widetilde X$ in Eqs.~(\ref{eq:d}) and (\ref{eq:k}). Somewhat
stronger constraints apply from CP violation in the neutral $D$ and
$K$ systems, if one assume phases of order one in the relevant scalar
couplings.

The constraint (\ref{eq:d}) is relevant, in principle, to cases II, IV
and (with the above mentioned modification) VI. The constraint
(\ref{eq:k}) is relevant, in
principle, to cases I, III and (with the above mentioned modification) V.

For cases I and V, the contribution to $K^0-\overline{K}{}^0$ is
suppressed by the small CKM combination $(V_{td}V_{ts}^*)^2$.
Consequently, these cases remain unconstrained in the region of
parameter space relevant for the forward-backward asymmetry. For cases
II and VI, the contribution to $D^0-\overline{D}{}^0$ is suppressed by
the small CKM combination $(V_{ub}V_{cb}^*)^2$. Consequently, also
these cases remain unconstrained in the region of parameter space
relevant for the forward-backward asymmetry. These statements are
valid even when taking into account the constraints from CP violation
in the $K$ and $D$ systems.

For case III, the constraint from $K^0-\overline{K}{}^0$ mixing reads
\beq\label{eq:kiii}
|\lambda|^4\mathcal{F}\left(\frac{m_t^2}{M^2}\right)<
4\times10^{-4}\left(\frac{M}{250\,\rm GeV}\right)^{2},
\eeq
thus ruling out case III from explaining the $A_h^{t\bar t}$. (To the
best of our understanding, the model of Ref.~\cite{Nelson:2011us}
violates the constraint (\ref{eq:kiii}) by two orders of magnitude and
is therefore excluded.)

For case IV, the constraint from $D^0-\overline{D}{}^0$ mixing reads
\beq
|\lambda|^4\mathcal{F}\left(\frac{m_t^2}{M^2}\right)<2.7\times10^{-4}
\left(\frac{M}{250\,\rm GeV}\right)^{2},
\eeq
thus ruling out case IV from explaining the $A_h^{t\bar t}$.

\subsection{Anomalous $B$ decays}
\label{ssec:bdec}
Anomalous $b$ decays proceed via tree level exchange of the charged
scalar, with
\beq\label{eq:Heffb}
\mathcal{H}_{\rm eff}\supset-\frac{4}{M^2}\left(X^\dag
  V\right)_{ji}^*\,\left(X^\dag V\right)_{k3}\,\left(d_{Li}^\dag
  u_{Rj}\right)\left(u_{Rk}^\dag b_L\right)+{\rm h.c.},
\eeq
or via exchange of the neutral scalar, with
\beq\label{eq:tilHeffb}
\mathcal{H}_{\rm eff}\supset-\frac{4}{M^2}\left(\tilde X^\dag
  V\right)_{ji}^*\,\left(\tilde X^\dag V\right)_{k3}\,\left(d_{Li}^\dag
  d_{Rj}\right)\left(d_{Rk}^\dag b_L\right)+{\rm h.c.},
\eeq
where brackets denote SU(3)$_c$ singlets. Of particular interest are
the charmless decays, $b\to u\bar ud$, $b\to u\bar us$, $b\to d\bar
dd$ and $b\to d\bar ds$.

Among the four limiting cases remaining to consider [cases I
(\ref{eq:x31}), II (\ref{eq:vx31}), V (\ref{eq:tilx31}) and VI
(\ref{eq:tilvx31})] only in cases I and V there is a $\phi$-mediated
tree level contribution to the decays considered here. In cases II and
VI the scalar component coupling to the bottom quark ($\phi^-\bar b_L
u_R$ or $\phi^{0*}\bar b_L d_R$) has no additional coupling to quarks,
and consequently these cases are unconstrained by $B$ decays.

We calculate the scalar-mediated contributions to $B$-meson decays
into the seven final states, $\pi^+\pi^-$, $\pi^-\pi^0$, $\pi^0\pi^0$,
$\pi^-\overline K^0$, $\pi^0 K^-$, $\pi^+ K^-$ and $\pi^0\overline
K^0$. We use the operator basis as defined in
Ref.~\cite{Beneke:2001ev}. At the scale $M$, the effective Hamiltonian
terms (\ref{eq:Heffb}) and (\ref{eq:tilHeffb}) correspond to the
$\mathcal{O}^{(i)}_6$ and $\mathcal{O}^{(i)}_8$ operators,
\beq\label{eq:Ob}
\mathcal{O}^{(i)}_6=4\sum_{q=u,d}\left(\bar b_L^\alpha \gamma^\mu
  d_{Li\beta}\right)
\left(\bar q_R^\beta \gamma_\mu q_{R\alpha}\right),\;\;
\mathcal{O}^{(i)}_8=4\sum_{q=u,d}\frac{3e_q}{2}
\left(\bar b_L^\alpha \gamma^\mu d_{Li\beta}\right)
\left(\bar q_R^\beta \gamma_\mu q_{R\alpha}\right),
\eeq
where ${\cal H}_{eff} \supset \sum_i \left(z^{(i)}_6 {\cal
    O}^{(i)}_6+z^{(i)}_8 {\cal O}^{(i)}_8\right)$. In case I, these
operators are generated with coefficients
\begin{subequations}\begin{align}
\delta z^{(i)}_6(M)
&\approx\frac{G_F}{3\sqrt{2}}\left(\frac{M}{250\,\rm
    GeV}\right)^{-2} \left(X^\dag V\right)_{1i}\,\left(X^\dag V\right)^*_{13},\\
\delta z^{(i)}_8(M)
&\approx\frac{2G_F}{3\sqrt{2}}\left(\frac{M}{250\,\rm
    GeV}\right)^{-2} \left(X^\dag V\right)_{1i}\,\left(X^\dag V\right)^*_{13}.
\end{align}\end{subequations}
In case V, these operators are generated with coefficients
\begin{subequations}\begin{align}
\delta z^{(i)}_6(M)
&\approx\frac{2G_F}{3\sqrt{2}}\left(\frac{M}{250\,\rm
    GeV}\right)^{-2} \left(\tilde X^\dag V\right)_{1i}\,\left(\tilde X^\dag V\right)^*_{13},\\
\delta z^{(i)}_8(M)
&\approx-\frac{2G_F}{3\sqrt{2}}\left(\frac{M}{250\,\rm
    GeV}\right)^{-2} \left(\tilde X^\dag V\right)_{1i}\,\left(\tilde X^\dag V\right)^*_{13}.
\end{align}\end{subequations}
QCD running from the scale $M$ down to the scale $m_B$ is captured at
leading log approximation (LLA)~\cite{Buchalla:1995vs} by ($i=d,s$)
\beq
\delta z^{(i)}_6(m_B)\approx1.5\,\delta z^{(i)}_6(M),\;\;
\delta z^{(i)}_8(m_B)\approx1.7\,\delta z^{(i)}_8(M).
\eeq
In addition, operator mixing at the ten percent level is induced, and
is taken into account in our numerical calculations. We compute the
relevant branching ratios using QCD
factorization~\cite{Beneke:1999br}.

Comparing the flavor factor in the $\phi$-mediated diagrams to that in
the $W$-mediated tree level diagrams, we observe that the former is
mildly enhanced in $b\to u\bar ud$, but strongly enhanced in $b\to
u\bar us$:
\beq\label{eq:flarat}
\left|\frac{V_{tb}V_{td}}{V_{ub}V_{ud}}\right|\sim2.5,\ \ \
\left|\frac{V_{tb}V_{ts}}{V_{ub}V_{us}}\right|\sim50.
\eeq
In Table~\ref{tab:bdec} we provide a full list of the branching ratios
for the relevant $B$ decays. For each process, we present the
experimental value, taken from Ref.~\cite{PDG}, and the
scalar-mediated contributions calculated for $M=250$ GeV and
$|\lambda|=1$.
\begin{table}[t]
  \caption{Branching ratios for the relevant charmless $B$ decays. For
the experimental result, we provide the central value~\cite{PDG}.
The experimental errors are at most of order $10\%$. For case I,
the results scale with $|\lambda|^4 \left(\frac{M}{250\,{\rm
      GeV}}\right)^{-4}$,
where $M$ is the mass of the charged scalar. For case V,
the results scale with $|\widetilde\lambda|^4\left(\frac{M}{250\,{\rm
      GeV}}\right)^{-4}$,
where $M$ is the mass of the neutral scalar.}
\label{tab:bdec}
\begin{center}
\begin{tabular}{c|cccc} \hline\hline
  \rule{0pt}{1.2em}%
 Branching ratio & Exp. value & case I & case V \cr \hline
 $\overline B^0\to\pi^+\pi^-$ &$5.1\times10^{-6}$ & $2.0\times10^{-4}$& $6.9\times10^{-7}$\cr
$B^-\to\pi^-\pi^0$ & $5.7\times10^{-6}$ & $5.9\times10^{-5}$& $5.9\times10^{-5}$ \cr
$\overline B^0\to\pi^0\pi^0$& $1.6\times10^{-6}$ & $6.5\times10^{-6}$ & $4.8\times10^{-5}$ \cr
$B^-\to\pi^-\overline K^0$ &$2.3\times10^{-5}$ & $7.4\times10^{-6}$ & $4.7\times10^{-3}$ \cr
$B^-\to\pi^0 K^-$ &$1.3\times10^{-5}$ & $1.2\times10^{-3}$ & $1.5\times10^{-4}$ \cr
$\overline B^0\to\pi^+ K^-$ & $1.9\times10^{-5}$ & $4.6\times10^{-3}$ & $1.6\times10^{-5}$ \cr
$\overline B^0\to\pi^0 \overline K^0$ & $9.5\times10^{-6}$ & $1.7\times10^{-4}$ & $1.0\times10^{-3}$ \cr
\hline\hline
\end{tabular}
\end{center}
\end{table}

In case I, in all modes except $B^-\to \overline K^0\pi^-$, the
$\phi$-mediated contribution, with $|\lambda|^2\left(\frac{M}{250\,\rm
    GeV}\right)^{-2}=\mathcal{O}(1)$, as required to explain
$A_h^{t\bar t}$, is significantly larger than the experimental value.
The strongest enhancement applies to ${\rm BR}(\overline
B^0\to\pi^+ K^-)$, where the scalar contribution is a factor of about
240 above experimental bounds. The flavor ratio in Eq.
(\ref{eq:flarat}) provides an enhancement of about three orders of
magnitude, but the heavier scalar mass sets off part of this
enhancement.
(To the best of our understanding, the model of Ref.
\cite{Nelson:2011us} enhances the $b\rightarrow u \bar u s$
transitions by about two orders of magnitude and is therefore
excluded. This point was made in Ref.~\cite{Zhu:2011ww} which
considers, however, the branching ratio for $B^+\to\pi^+K^0$. This
decay is a $b\rightarrow d \bar d s$ transition that is only generated
by RGE effects. Thus this channel puts only mild constraints on the
model, in comparison to $b\to u\bar us$ transitions.)

In case V, in all modes except $\overline{B}^0\to \pi^+ \pi^-$ and
$\overline{B}^0\to \pi^+ K^-$, the scalar-mediated contribution, with
$|\widetilde\lambda|^2\left(\frac{M}{250\,\rm
    GeV}\right)^{-2}=\mathcal{O}(1)$, as required to explain
$A_h^{t\bar t}$, is significantly larger than the experimental value.
The strongest enhancement applies to ${\rm BR}(B^-\to \overline K^0 \pi^-)$,
where the scalar contribution is enhanced by a factor of about 200
above the experimental bound.

We conclude that charmless $B$ decays exclude the possibility that
$A_h^{t\bar t}$ is accounted for by a weak doublet scalar with
couplings of type I or V.  We thus find that only cases II and VI
survive the constraints from flavor changing processes and can provide
a viable mechanism for $A_h^{t\bar t}$.

In principle, we can consider the scalar couplings of cases II and
VI simultaneously. The relevant part of the Lagrangian is then:
\beqa\label{eq:lagyrdr}
 {\cal L}&\supset&2\lambda q_{L3}^\dagger\Phi u_R+\tilde\lambda
 q_{L3}^\dagger\tilde \Phi d_R+{\rm h.c.}\\
 &=&2\lambda(u_{Li}^\dagger V_{ib}\phi^0 u_R+b_L^\dagger \phi^- u_R)+
 2\tilde\lambda(u_{Li}^\dagger V_{ib}\phi^+ d_R-b_L^\dagger
 \phi^{0*} d_R) +{\rm h.c.}\no
\eeqa
Integrating out the scalar field, we get (among others) the following
interesting four-quark terms:
\beq\label{eq:leff}
{\cal L}_{\rm eff}=\frac{\lambda\tilde\lambda
  V_{ib}}{m_+^2}(b_L^\dagger u_R)(u_{Li}^\dagger d_R)-
\frac{\lambda\tilde\lambda
  V_{ib}}{m_0^2}(b_L^\dagger d_R)(u_{Li}^\dagger u_R),
\eeq
where $m_0$ and $m_+$ denote the masses of, respectively, the neutral
and charged scalars.  These terms lead, in particular, to $b\to u\bar
cd$ decays, with CKM suppression by a factor of $V_{cb}\sim0.04$. In
contrast, within the SM, the $W$-mediated diagram is CKM suppressed by
$V_{ub}V_{cd}\sim0.0008$, a factor of 50 smaller.

The hadronic modes that are described by this ``wrong sign'' quark
transition are $B^+\to D^+\pi^0$, $B^+\to D^0\pi^+$, $B^0\to
D^0\pi^0$, $B^0\to D^+\pi^-$ and the analogous modes with $D\to
D^*$ and/or with $\pi\to\rho$. (Of course, charge-conjugate modes are
implied.)

In the PDG~\cite{PDG}, one finds only a single attempt to measure one
of these modes\footnote{Note that the range quoted, ${\rm BR}(B^0\to
  D^{+}\pi^-)=(4.6\pm0.4)\times10^{-5}$, is {\it not} an experimental
  measurement, but rather a theoretical calculation based on a
  measurement of the $D_s^+\pi^-$ mode \cite{:2010be}.} \cite{:2008av}
\beq\label{eq:exp}
{\rm BR}(B^+\to D^{*+}\pi^0)<3.6\times10^{-6}.
\eeq
The SM expectation, based on extracting the parameter $r$,
\beq\label{eq:rsm}
r\equiv\sqrt{\frac{\tau_0}{\tau_+}\frac{2\,{\rm BR}(B^+\to
    D^{*+}\pi^0)}{{\rm BR}(B^0\to D^{*-}\pi^+)}},
\eeq
from isospin relation and two measured branching fractions,
\beq
r=\tan\theta_c\frac{f_{D^*}}{f_{D_s^*}}\sqrt{\frac{{\rm BR}(B^0\to
    D_s^{*+}\pi^-)}{{\rm BR}(B^0\to D^{*-}\pi^+)}}\approx0.02,
\eeq
and on $\tau_+/\tau_0=1.071\pm0.009$ and ${\rm BR}(B^0\to
D^{*-}\pi^+)=(2.76\pm0.21)\times10^{-3}$, is
\beq\label{eq:sm}
{\rm BR}(B^+\to D^{*+}\pi^0)^{\rm SM}\approx5.9\times10^{-7}.
\eeq
Additional support that the $b\to u\bar cd$ transitions are doubly
Cabibbo suppressed (DCS) can be found in \cite{Ronga:2006hv},
reporting on measuring CP violation in the Cabibbo favored (CF) modes
$B^0\to D^{*-}\pi^+$ and $B^0\to D^-\pi^+$. In their Fig. 11, one can
read an upper bound of order a few percent on the ratio between the
DCS and CF amplitudes for each of the two cases.

Comparing the experimental bound (\ref{eq:exp}) to the SM prediction
(\ref{eq:sm}), we conclude that the rate is enhanced by no more than a
factor of order 6, while Eq. (\ref{eq:leff}) predicts enhancement by
about two-to-three orders of magnitude,
\beq\label{eq:scalar}
\frac{\Gamma(B^+\to D^{*+}\pi^0)^{\Phi}}
{\Gamma(B^+\to D^{*+}\pi^0)^{\rm SM}}&\approx&\frac{|\lambda \tilde\lambda|^2}{g^4}
\left|\frac{V_{cb}}{V_{ub}V_{cd}}\right|^2
\left(\frac{m_W}{m_0}\right)^4
\left(\frac{\langle D^{*+}\pi^0|(b_L^\dagger d_R)(c_{L}^\dagger u_R)|B^0\rangle}
{\langle D^{*+}\pi^0|(b_L^\dagger\gamma_\mu u_L)(c_{L}^\dagger
  \gamma^\mu d_L)|B^0\rangle}\right)^2\no\\
  &\sim&2500\ \frac{|\lambda \tilde\lambda|^2}{g^4}\left(\frac{m_W}{m_0}\right)^4.
\eeq
Thus, simultaneous order one couplings to both the up-quark (case II)
and down-quark (case VI) are excluded.

In Section \ref{sec:top} we concluded that, in order to account for
$A_h^{t\bar t}$, the scalar doublet should have order one couplings to
the top and a first generation quark. In this section we found that
coupling to the first generation quark doublet should be avoided. The
order one coupling can be to either $u_R$ or $d_R$, but not to both.
In the language of our six limiting cases, only one of the two types
II or VI is allowed by flavor constraints.

\section{Electroweak constraints}
\label{sec:ewpm}
%

\subsection{The $S$ and $T$ parameters}
\label{ssec:t}
The weak doublet can contribute to the electroweak $S$ and $T$
parameters. The PDG constraints read~\cite{PDG}
\beq
S&=&0.01 \pm 0.10(-0.08)\,,\\
T&=&0.03\pm 0.11(+0.09)\,.
\eeq
The central values corresponds to $m_h=117$~GeV, while the shifts in
parenthesis corresponds to $m_h = 300$~GeV.

We use the expressions for $T$ and $S$ in two Higgs doublet models
from Ref.~\cite{Branco:2011iw}. Taking the approximation of non-mixed
neutral scalars, where one is degenerate with $m_h$ and the others
have a common mass $m_0$, and denoting the mass of the charged
component by $m_+$, we have
\beq
T&=&\frac{1}{8\pi s_W^2 m_W^2}\left(\frac{m_0^2+m_+^2}{2}
  -\frac{m_0^2 m_+^2}{m_+^2-m_0^2} \ln\frac{m_+^2}{m_0^2}\right) \, ,\\
S&=& \frac{1}{24 \pi}\left[(s_W^2-c_W^2)^2 G\left(z_+,z_+\right)
  +G\left(z_0,z_0\right)\right]\, ,
\eeq
where
\beq
z_a&\equiv& m_a^2/m_Z^2,\\
G(x,y)&=&-\frac{16}{3}+5(x+y)-2(x-y)^2 + 3\left[\frac{x^2+y^2}{x-y}
  -x^2+y^2 +\frac{(x-y)^3}{3}\right]\ln\frac{x}{y}\no \\
& &  +\left[1-2(x+y)+(x-y)^2\right]f(x+y-1,1-2(x+y)+(x-y)^2), \\
f(z,w)&=&\left\{ \begin{array}{ll}
\sqrt{w} \ln \left|\frac{z-\sqrt w}{z+\sqrt w}\right| & w>0,\no\\
0 & w=0\no, \\
2\sqrt{-w} \arctan\frac{\sqrt{-w}}{z} & w<0.\\
\end{array} \right.
\eeq

The $S$ parameter provides no meaningful constraint on the parameters
of our model. The $T$ parameter, on the other hand, constrains the
mass splitting within the scalar doublet. The $2\sigma$ range for $T$
gives a conservative bound on the maximal allowed mass splitting:
\beq\label{eq:tcons}
\left|\frac{m_+-m_0}{M}\right|\lsim0.45\ \frac{250\
  {\rm GeV}}{M},
\eeq
where $M=\frac12(m_++m_0)$.  For case II (VI), $A_h^{t\bar t}$
restricts the mass $m_0$ ($m_+$). Then, the $T$ constraint of
Eq.~(\ref{eq:tcons}) allows the mass $m_+$ ($m_0$) to be shifted from
$m_0$ ($m_+$) by at most $\sim110$~GeV. These results are consistent with~\cite{Cao:2011yt}.

Explaining $A^{t\bar t}$ requires the mass of the relevant weak-doublet component to be in the range $100-130$~GeV. Thus, we can now justify neglecting the mass splitting between the
weak-doublet components in the discussion of flavor constraints in
Section \ref{sec:flavor} and the use of the benchmark value of $250$~GeV:
\begin{itemize}
\item In cases III and IV, $A^{t\bar t}$ is generated by the exchange
  of both the neutral and the charged scalars, and so neglecting the mass
  splitting in the meson-mixing analysis, which led to their
  exclusion, is justified.
\item In cases I and V, $A^{t\bar t}$ is generated by the exchange of
  one component while meson-mixing and $B$ decays are sensitive to the mass of the
  other scalar component. There, the $T$-parameter serves to restrict
  the splitting so that the analysis is valid.
\item In cases II and VI, meson-mixing is generated by the same scalar
  component responsible for $A^{t\bar t}$ and there is no ambiguity.
  In the case where couplings of both type II and VI exist, the mixed $B$ decays are sensitive to the masses of both scalar components.
  The use of a single benchmark mass value in the analysis is valid and the net result is unaltered.
\end{itemize}

\subsection{$R_b$}
Contributions to $Z\to b_L\bar b_L$ can arise from $(V^\dag X)_{3i}$
or $(V^\dag \widetilde X)_{3i}$ ($i=1,2,3$) couplings via one loop
diagrams with an internal quark and scalar. In cases II and VI, only
the $i=1$ terms exist. Explicitly, for case II the relevant coupling
is $\lambda=(V^\dag X)_{31}$ and the internal quark is $u_R$, while in
case VI the relevant coupling is $\widetilde\lambda=(V^\dag \widetilde
X)_{31}$ and the internal quark is $d_R$.

We denote the intermediate quark masses by $m_u$ and $m_d$ and the
tree level $Z$ couplings to the different particles by
$g_{d_L},\;g_{u_R},\;g_{d_R},\;g_{\phi^-}$ and $g_{\phi^0}$. We
neglect terms of $\mathcal{O}\left({m_b}/{M}\right)$ and use
$g_{u_R}+g_{\phi^-}-g_{d_L}=0$ and $g_{d_R}-g_{\phi^0}-g_{d_L}=0$.

To leading order in $|\lambda|^2/(4\pi)^2$, we obtain the effective
shift to the $Z_\mu\bar b_L\gamma^\mu b_L$ vertex for case II:
\beq\label{eq:dgg}
\!\!\!\!\!\!\!\!\!\!\left.\frac{\delta g_{d_L}}{g_{d_L}}\right|_{\rm
  II}&=&
\frac{4|\lambda|^2}{(4\pi)^2}\mathcal{F'}(r_Z,r_u),\\
\!\!\!\!\!\!\!\!\!\!\mathcal{F'}(r_Z,r_u)&=&
\int dxx\ln\left(\frac{\Delta_c}{M^2}\right)\no\\
&-&\int dx dydz\delta(x+y+z-1)\left[\frac{g_{u_R}}{g_{d_L}}
  \left(\ln\left(\left|\frac{\Delta_a}{M^2}\right|\right)+
    \frac{zyr_Z}{\Delta_a}+1\right)+\frac{g_{\phi^-}}{g_{d_L}}
  \ln\left(\left|\frac{\Delta_b}{M^2}\right|\right)\right].\no
\eeq
with
\beq&\frac{\Delta_c}{M^2}=x+r_u(1-x),\ \ \
\frac{\Delta_a}{M^2}=\frac{\Delta_c}{M^2}-r_Zzy,\ \ \
\Delta_b=\Delta_a(x\leftrightarrow 1-x),\\
&r_u={m_u^2}/{M^2},\ \ \
r_Z={m_Z^2}/{M^2}.\no
\eeq
For case VI, we obtain
\beq\label{eq:dggsix}
\left.\frac{\delta g_{d_L}}{g_{d_L}}\right|_{\rm
  VI}=\left.\frac{\delta g_{d_L}}{g_{d_L}}\right|_{\rm
  II}(\lambda \to \widetilde \lambda,\ m_u\to m_d,\ g_{u_R}
\to g_{d_R},\ g_{\phi^-}\to -g_{\phi^0}).
\eeq

In addition to the shift Eq.~(\ref{eq:dgg}) (or (\ref{eq:dggsix})), the
new scalars introduce tensor as well as imaginary vector terms to the
$Z\to b_L\bar b_L$ amplitude. Since these other terms do not interfere
with the leading SM diagram, they contribute only at next order in
$\frac{|\lambda|^2}{(4\pi)^2}$ (or
$\frac{|\widetilde\lambda|^2}{(4\pi)^2}$) and we omit them here.

The shift in the coupling induces a shift in $R_b$ according to
\beq
\frac{\delta R_b}{R_b}\approx1.5\ \frac{\delta g_{d_L}}{g_{d_L}}.
\eeq
The experimental $1\sigma$ bound is $\delta R_b/R_b\lsim
0.003$~\cite{PDG} which is satisfied in our model for $M\gsim70$ GeV.
Thus, our model parameters in both cases II and VI are unconstrained
by $R_b$ in the relevant parameter space.

\section{Additional collider constraints}
\label{sec:topsss}
%
\subsection{Single top production}
D0 have performed a model-independent measurement of the $tbq$
production cross section, where $q$ is a light
quark~\cite{Abazov:2011rz}:
\beq\label{eq:ttev}
\sigma(p\bar p\rightarrow tbq+X)=2.90\pm 0.59\ {\rm pb},
\eeq
in good agreement with the SM $t-$channel $tbq$ result of
$2.26\pm0.12$~pb~\cite{Kidonakis:2006bu}.

In the cases II and VI, single tops can be produced by the process $qg\to t\phi$, where $q=u,d$. We find
the matrix element for such a process to be given by
\beq
\overline{|{\cal M}|^2}_{gu\to t\phi^0}=-\frac{|\lambda|^2 g_s^2}{3}
\left[\frac{\tilde s}{\tilde u}+\frac{\tilde u}{\tilde s}
  +\frac{2(\tilde t+\tilde s)(\tilde t + \tilde u)}{\tilde u \tilde s}
  -\frac{2m_t^2}{\tilde u^2}(M^2-m_t^2)\right],
\eeq
where $M=m_0$. For the process $gd\to t\phi^-$ the replacement
$\lambda\to\widetilde \lambda$ should be made, and $M=m_+$. The parton
level Mandelstam variables obey $\tilde u+ \tilde t+\tilde s=M^2-m_t^2$.

Note, however, that in case II, $\phi^0$ that is produced together
with the top quark will decay into an up-sector quark and an up-sector
antiquark. In case VI, $\phi^-$ that is produced with the top quark,
will decay into the down quark and an up-sector antiquark. In either
case, the scalar doublet will not decay into a bottom (anti)quark.

In principle, such a process might still be constrained by the $tbq$
measurement, since the data used in the analysis included a singly
$b$-tagged sample as well. We find that the cross section for
the production of $t\phi$ is sizable, of order
3 pb in the allowed parameter space for $A^{t\bar t}$. Single top production can then become competitive with the
other top-related constraints ($t\bar t$ cross section and $A_l^{t\bar t}$).

Additionally, the CMS collaboration has recently reported a similar measurement at the LHC~\cite{Chatrchyan:2011vp}:
\beq\label{eq:tlhc}
\sigma_t=84\pm30\ {\rm pb},
\eeq
consistent with the SM $t-$channel result of $64.3\pm2.2$~pb~\cite{Kidonakis:2011wy}. The cross section of $t\phi$ at the LHC in the allowed parameter space for $A^{t\bar t}$ is again sizable, of order $140-170$~pb.

Given the complexity of the analyses in~\cite{Abazov:2011rz} and~\cite{Chatrchyan:2011vp}, and the different kinematics of single top production in our model compared to the SM, a direct comparison of the single top production cross section in our model to Eqs.~\eqref{eq:ttev} and~\eqref{eq:tlhc} is potentially misleading. We find that a dedicated study is required in order to establish the applicability of the single top measurements at the Tevatron and LHC to our model.

\subsection{Top decay}\label{sec:tdec}
If the extra scalar masses are light enough, new decay channels
open for the top:
\beqa
\Gamma(t\to\phi^0 u_i)&=&\frac{m_t}{8\pi}
\left(1-\frac{m_0^2}{m_t^2}\right)^2\left(|X_{i3}|^2+|X_{3i}|^2\right),\no\\
\Gamma(t\to\phi^+ d_i)&=&\frac{m_t}{8\pi}
\left(1-\frac{m_+^2}{m_t^2}\right)^2\left(|(V^\dagger
  X)_{i3}|^2+|\widetilde X_{3i}|^2\right).
\eeq
In cases II and VI we then have
\beqa
\left.\frac{\delta\Gamma_t}{\Gamma_t^{\rm SM}}\right|_{\rm II}&\approx &
5.2\left(1-\frac{m_0^2}{m_t^2}\right)^2|\lambda|^2,\no\\
\left.\frac{\delta\Gamma_t}{\Gamma_t^{\rm SM}}\right|_{\rm IV}&\approx &
5.2\left(1-\frac{m_+^2}{m_t^2}\right)^2|\widetilde\lambda|^2,
\eeq
using $\Gamma_t^{\rm SM}\approx1.3$~GeV. The direct
measurement of the top-quark width puts an upper bound of
$\Gamma_t<7.6$~GeV~\cite{Aaltonen:2010ea}. A model-dependent indirect
measurement of the total top width, from the partial decay width
$\Gamma(t \to Wb)$ measured using the t-channel cross section for
single top quark production and from the branching fraction ${\rm
  BR}(t \to Wb)$ gives
$\Gamma_t=1.99^{+0.69}_{-0.55}$~GeV \cite{Abazov:2010tm}. We find that
measurements of the top width do not constrain the parameter space of
our model relevant for producing the forward-backward asymmetry.

The above modification of the top width can effect measurements of the single top and $t\bar t$ production cross sections which typically assume SM $Wb$ final states.\footnote{We thank Pedro Schwaller for a comment on this point.} A sizable branching fraction of the top into three light quarks, as predicted in our scenario, might be translated into a reduction in the inferred cross section. This reduction effect might become comparable in magnitude to the scalar exchange contributions in some of the parameter space relevant for $A^{t\bar t}$ and cancellations may occur. A naive bound on the size of this effect can be obtained by assuming that the non-SM top decays completely evade the experimental analyses. In this limit, we find that the allowed parameter space shifts to larger values of $\lambda\gsim1$ for roughly the same masses $M\sim 100-130$~GeV.  The $t\phi$ production cross section is further enhanced in this naive estimate, and so clarifying the applicability of the single top measurements to our model becomes more urgent.

\subsection{Same sign tops}
Production of same sign tops at the Tevatron due to the couplings of
interest to us is negligible. The reason is that the same sign top
production is proportional to the product of the couplings of Eq.
(\ref{eq:x31}) [case I] and of Eq. (\ref{eq:vx31}) [case II]. However,
neutral meson-mixing requires that the couplings of Eq. (\ref{eq:x31})
are strongly suppressed.  Thus, production of same sign
tops~\cite{CDF:tt} does not constrain the parameter space of our
models. (The contribution from weak doublet scalar exchange to
  same sign tops was considered in
  Ref.~\cite{AguilarSaavedra:2011zy}. The large effects that they
  derive arise in the region that is excluded by flavor constraints.)

\subsection{Dijet constraints}
The weak doublet contributes to dijet production via both $t-$ and
$s-$channel exchange of $\phi^0$. There are no dijet constraints on
the model from the Tevatron, since for $M\sim 100-130$~GeV the
Tevatron has large SM dijet backgrounds~\cite{Aaltonen:2008dn}. In
principle, bounds could arise from measurements by the UA2 collaboration at
the CERN SPS collider~\cite{Alitti:1993pn}. However, the $t-$channel
exchange would not have been picked up by the UA2 search, and the
$s-$channel exchange is strongly suppressed by CKM factors, $\sim
V_{ub}^2$ (or $V_{cb}^2$,
paying the price of one $c$ quark pdf). We conclude that dijet
constraints from $p\bar p$ colliders play no role here.

As concerns the LHC, contributions to dijet production from
the $s-$ and $t-$channels of $gu\rightarrow q\Phi\rightarrow {\rm
  dijets }$ resemble non-dominant QCD backgrounds and are
not constrained by the dijet angular distribution studies presented by
CMS~\cite{Khachatryan:2011as} and ATLAS~\cite{Aad:2011aj}.

\section{Summary}
\label{sec:sum}
We investigated whether the large value reported by CDF for the
forward backward asymmetry in $t\bar t$ production at large invariant
mass $M_{t\bar t}$ can be accounted for by tree level scalar exchange.
We considered top-related measurements, flavor constraints, and
electroweak precision measurements. We reached the following
conclusions:
\begin{itemize}
\item Out of the eight possible scalar representations that are
  relevant to $A^{t\bar t}$, only the color-singlet weak-doublet
  $\Phi(1,2)_{-1/2}$ can enhance $A_h^{t\bar t}$ and remain
  consistent with the total and differential $t\bar t$ cross section.
  Roughly speaking, the relevant Yukawa coupling should be $\mathcal{O}(1)$,
  and the mass of the scalar should be below $\sim 130$~GeV.
\item Two types of couplings of $\Phi$ can contribute to $u\bar u \to
  t\bar t$: $X_{13} q^\dagger_{L1} \Phi t_R$ and $X_{31}
  q^\dagger_{L3} \Phi u_R$. There is no tension with the differential
  or total $t\bar t$ production cross section. Both couplings are
  constrained by flavor physics:
\begin{enumerate}
\item The $X_{13}$ coupling is strongly constrained by
  $K^0-\overline{K}{}^0$ and/or $D^0-\overline{D}{}^0$ mixing, and so
  cannot generate a large $A^{t\bar t}$.
\item The $X_{31}$ coupling is not strongly constrained by neutral
  meson mixing, or by $R_b$. If $\phi^-$ couples to the three
  left-handed down generations with CKM-like suppression ${\cal O}(V_{tq})$, then
  it contributes to the branching ratio of $\overline{B^0}\rightarrow\pi^+
  K^-$ more than two orders of magnitude above the experimental
  bounds. If, on the other hand, the $X_{31}$ coupling is carefully aligned so
  that $\phi^-$ couples only to $b_L$ (but not to $s_L$ and $d_L$),
  then it can be large enough to explain $A^{t\bar t}$.
\end{enumerate}
\item $\Phi$ could also affect $A^{t\bar t}$ by mediating $d\bar d\to
  t\bar t$ with coupling $\widetilde X_{31} q^\dagger_{L3} \Phi d_R$.
  The coupling would need to be bigger than the $X_{31}$ coupling to
  overcome the small $d\bar d$ luminosity at the Tevatron. This is in tension with the $t\bar t$ production cross section.
  Again, flavor constraints are very
  restrictive:
\begin{enumerate}
\item Similarly to the case of $X_{31}$, there is no strong constraint
  from either neutral meson mixing or $R_b$. If $\phi^0$ couples to
  the three left-handed down generations with CKM-like suppression
  ${\cal O}(V_{tq})$, then it contributes to the branching ratio of
  $B^-\rightarrow \pi^- \overline{K^0}$ more than two orders of
  magnitude above the experimental bound. If, on the other hand, the
  $\widetilde X_{31}$ coupling is carefully aligned so that $\phi^0$
  couples only to $b_L$ (but not to $s_L$ and $d_L$), then it can be
  large enough to explain $A^{t\bar t}$.
\item The $X_{31}$ and $\widetilde X_{31}$ couplings cannot be
  simultaneously order one, because then the upper bound on the
  branching ratio of $B^+\to D^{*+}\pi^0$ is violated by more than two
  orders of magnitude.
\end{enumerate}
Thus, the coupling $\tilde{X}$ cannot play a significant role in explaining $A^{t\bar t}$.
\item The flavor constraints that we derive might be circumvented if
  the contributions to flavor changing processes cancel against
  contributions from additional scalar doublets. For this to happen,
  special relations between the couplings of the various scalars must
  apply. Such relations might appear in models of minimal flavor
  violation.  An example can be found in Ref. \cite{Babu:2011yw}. (To
  fully satisfy the flavor constraints, degeneracy constraints on
  the scalar spectrum of this model should hold.)
\item The new physics contribution to single top production at the Tevatron and LHC is comparable to or larger than the electroweak SM single top production; However the event topology is different. The sensitivity of existing experimental searches, designed to extract SM-like event topologies, to single top production in the weak-doublet model is hard to assess, and a dedicated study is required.
\item No sizable distortion of the differential or inclusive $t\bar t$
production cross section is expected at the LHC.
\end{itemize}

We conclude that the interplay between collider physics and flavor
physics singles out a weak-scale color-singlet weak-doublet scalar,
with a very non-generic flavor structure of Yukawa couplings, as the
only viable candidate among the scalars to account for a large forward
backward asymmetry in $t\bar t$ production.

\begin{appendix}

\section{Calculating the $t\bar t$ production cross section}
\label{app:ope}
The details of the calculation of the $t\bar t$ production cross
section are collected here.  We use the following (over-complete)
Lorentz basis for the relevant four-fermi flavor conserving operators:
\begin{subequations}
\label{eq:4f}\begin{align}\!\!\!\!\!
\mathcal{O}_{V_qV_t}&=\bar q\gamma^\mu q\,\bar t\gamma_\mu t,&
\mathcal{O}_{A_qA_t}&=\bar q\gamma^\mu\gamma^5 q\,\bar t\gamma_\mu\gamma^5 t,&
\mathcal{O}_{A_qV_t}&=\bar q\gamma^\mu\gamma^5 q\,\bar t\gamma_\mu t,&
\mathcal{O}_{V_qA_t}&=\bar q\gamma^\mu q\,\bar t\gamma_\mu\gamma^5 t,&\\
\mathcal{O}_{S_qS_t}&=\bar q q\,\bar t t,&
\mathcal{O}_{P_qP_t}&=\bar q\gamma^5 q\,\bar t\gamma^5 t,&
\mathcal{O}_{P_qS_t}&=i\,\bar q\gamma^5 q\,\bar t t,&
\mathcal{O}_{S_qP_t}&=i\,\bar q q\,\bar t\gamma^5 t,&\\
\mathcal{O}_{T_qT_t}&=\bar q\sigma^{\mu\nu} q\,\bar t\sigma^{\mu\nu} t,&
\mathcal{O}_{T'_qT'_t}&=\bar q\sigma^{\mu\nu}\gamma^5 q\,\bar t\sigma^{\mu\nu}\gamma^5 t,&
\mathcal{O}_{T'_qT_t}&=\,i\bar q\sigma^{\mu\nu}\gamma^5 q\,\bar t\sigma^{\mu\nu} t,&
\mathcal{O}_{T_qT'_t}&=\,i\bar q\sigma^{\mu\nu} q\,\bar t\sigma^{\mu\nu}\gamma^5 t,&
\end{align}
\end{subequations}
with $\sigma^{\mu\nu}=\frac{i}{2}\left[\gamma^\mu,\gamma^\nu\right]$.
We work with the signature of Peskin and Schroeder~\cite{Peskin:1995ev}.

For color contraction, we use the singlet and octet projections as our
basis and define:
\beq\label{eq:colbasis}
\mathcal{O}_{XY}^{8}=\left(T^a\right)_i^j\left(T^a\right)_k^l
\left(\bar q^i\Gamma_{Xq} q_j\right)\left(\bar t^k\Gamma_{Yt}
  t_l\right),\;\;
\mathcal{O}_{XY}^{1}=\frac{\sqrt{2}}{3}\delta_i^j\delta_k^l
\left(\bar q^i\Gamma_{Xq} q_j\right)\left(\bar t^k
  \Gamma_{Yt}t_l\right).
\eeq
Then, the coefficients of the four-quark operators are given by
\beq
{\cal L}_{{\rm eff},XY}=\frac{1}{M^2}\left(c_{XY}^{8}{\cal O}_{XY}^8+c_{XY}^1{\cal O}_{XY}^1\right).
\eeq
The basis Eq.~(\ref{eq:4f}) does not respect $SU(2)_L$. It is useful
because interference with the SM is proportional to $c^{8}_{V_qV_t}$
and $c^{8}_{A_qA_t}$.

The following Fierz identities are useful to our analysis:
\begin{subequations}
\begin{align}
\left(\psi_{2R}\psi_{4R}\right)\left(\psi_{1R}^\dag\psi_{3R}^\dag\right)&
=\frac{1}{2}\left(\bar\Psi_1\gamma^\mu P_{R}\Psi_2\right)
\left(\bar\Psi_3\gamma_\mu P_{R}\Psi_4\right),\\
\left(\psi_{1L}^\dag\psi_{4R}\right)\left(\psi_{3R}^\dag\psi_{2L}\right)&=
-\frac{1}{2}\left(\bar\Psi_1\gamma^\mu
  P_L\Psi_2\right)\left(\bar\Psi_3\gamma_\mu
  P_R\Psi_4\right).
\end{align}
\end{subequations}
Similar identities hold with $L\leftrightarrow R$.

We next write the relevant Lagrangian terms for each of the eight scalar
representations listed in Eq.~(\ref{eq:scarep}), and then the
effective four-fermi operators contributing to $t\bar t$ production
generated by integrating out the scalar field.  The effective
Lagrangian defined in this way is useful also in the case of light new
scalars.  The key point here is that only a single diagram contributes
to $t\bar t$ production cross section for a given representation. Thus
it is straightforward to extend the effective Lagrangian to include
the momentum dependence of the scalar propagator: One has to simply
replace $M^2\to M^2-q^2-iM\Gamma$. The anti-sextet and the triplet
contribute to $u\bar u\to t\bar t$ (or $d\bar d\to t\bar t$) via
$u$-channel exchange: $q^2=u=m_t^2-\frac{\tilde
  s}{2}\left(1+\beta_t\cos\theta\right)=m_t^2+\tilde u$. The octet and
the singlet contribute to $u\bar u\to t\bar t$ (or $d\bar d\to t\bar
t$) via $t$-channel exchange: $q^2=t=m_t^2-\frac{\tilde
  s}{2}\left(1-\beta_t\cos\theta\right)=m_t^2+\tilde t$.  We neglect
$SU(2)$-breaking mass splittings between the members of the scalar
multiplet.

\begin{itemize}
\item {\bf Color-sextet weak-singlet $\Phi\sim(\bar 6,1)_{-{4}/{3}}$}
\beq
\lag^{(\bar 6,1)_{-{4}/{3}}}=-M^2\Phi^{ij}\Phi_{ij}^\dag+\left[
2\sqrt{2}\lambda\Phi^{ij}
u_{Ri}t_{Rj}+{\rm h.c.}\right],
\eeq
\beq\label{eq:Leff6}
\lag_{\rm eff}^{\bar 6}
&=&\frac{|\lambda|^2}{M^2}
\left[\mathcal{O}^8_{V_uV_t}+\mathcal{O}^8_{A_uA_t}+\mathcal{O}^8_{A_uV_t}+\mathcal{O}^8_{V_uA_t}\right]\no\\
&+&\left(\mathcal{O}^8\to\sqrt{2}\,\mathcal{O}^1\right).
\eeq
\item {\bf Color-sextet weak-singlet $\Phi\sim(\bar 6,1)_{-{1}/{3}}$}
\beq
\lag^{(\bar 6,1)_{-{1}/{3}}}=-M^2\Phi^{ij}\Phi_{ij}^\dag+2\sqrt{2}\left[
\lambda_1\Phi^{ij}
d_{Ri}t_{Rj}+\lambda_2\Phi^{ij}\left(q_{Li}Q_{Lj}\right)+{\rm h.c.}\right],
\eeq
\beq\label{eq:Leff6}
\lag_{\rm eff}^{\bar 6}
&=&\frac{|\lambda_1|^2}{M^2}
\left[\mathcal{O}^8 _{V_dV_t}+\mathcal{O}^8 _{A_dA_t}+\mathcal{O}^8_{A_dV_t}+\mathcal{O}^8 _{V_dA_t}
\right]\no\\
&+&\frac{|\lambda_2|^2}{M^2}
\left[\mathcal{O}^8 _{V_dV_t}+\mathcal{O}^8 _{A_dA_t}
  -\mathcal{O}^8_{A_dV_t}-\mathcal{O}^8 _{V_dA_t}  \right]\no\\
&-&\frac{3\Re\left[\lambda_1\lambda^*_2\right]}{M^2}
\left[\mathcal{O}^8_{S_dS_t}+\mathcal{O}^8_{P_dP_t}
\right]-\frac{3\Im\left[\lambda_1\lambda^*_2\right]}{M^2}
\left[\mathcal{O}^8_{S_dP_t}+\mathcal{O}^8_{P_dS_t}
\right]\no\\
&-&\frac{\Re\left[\lambda_1\lambda^*_2\right]}{4M^2}
\left[\mathcal{O}^8_{T_dT_t}+\mathcal{O}^8_{T'_dT'_t}\right]-\frac{\Im\left[\lambda_1\lambda^*_2\right]}{4M^2}
\left[\mathcal{O}^8_{T_dT'_t}+\mathcal{O}^8_{T'_dT'_t}\right]\no\\
&+&\left(\mathcal{O}^8\to\sqrt{2}\,\mathcal{O}^1\right).
\eeq
The possibility of $\lambda_1\lambda_2=\mathcal{O}(1)$ makes the
numerical analysis of this model nontrivial. However, note that the
coupling $\lambda_2$ is subject to strong constraints from flavor
physics, including meson mixing and $b$ decays. We simplify the
analysis by imposing the conservative constraint
$|\lambda_2|^4\left(M/{\rm TeV}\right)^{-2}<10^{-2}$.
\item {\bf Color-sextet weak-triplet $\Phi\sim(\bar 6,3)_{-{1}/{3}}$}
\beq\label{eq:lagsextri}
\lag^{(\bar 6,3)_{-{1}/{3}}}&=&-M^2\Phi^{ij}\Phi_{ij}^\dag+\left[
2\sqrt{2}\lambda\Phi^{ij} \cdot\left(q_{Li}Q_{Lj}\right)+{\rm h.c.}\right],
\eeq
\beq
\lag_{\rm eff}^{\bar 6}&=&\frac{|\lambda|^2}{M^2}
\left[\mathcal{O}^8 _{V_uV_t}+\mathcal{O}^8 _{A_uA_t}
  -\mathcal{O}^8_{A_uV_t}-\mathcal{O}^8 _{V_uA_t} \right]
+\frac{1}{2}\times\left(u\leftrightarrow d\right)\no\\
&+&\left(\mathcal{O}^8\to\sqrt{2}\,\mathcal{O}^1\right).\
\eeq
\item {\bf Color-triplet weak-singlet $\Phi\sim(3,1)_{-{4}/{3}}$}
\beq
\lag^{(3,1)_{-{4}/{3}}}=-M^2\Phi_i\Phi^{\dag i}+
\left[2\lambda\epsilon^{ijk}\Phi_i
u_{Rj}t_{Rk}+{\rm h.c.}\right],
\eeq
\beq
\lag_{\rm eff}^{3}&=&-\frac{|\lambda|^2}{M^2}
\left[\mathcal{O}^8 _{V_uV_t}+\mathcal{O}^8 _{A_uA_t}
  +\mathcal{O}^8_{A_uV_t}+\mathcal{O}^8 _{V_uA_t}\right]\no\\
&+&\left(\mathcal{O}^8\to-\frac{1}{\sqrt{2}}\,\mathcal{O}^1\right).
\eeq
\item {\bf Color-triplet weak-singlet $\Phi\sim(3,1)_{-{1}/{3}}$}
\beq
\lag^{(3,1)_{-{1}/{3}}}=-M^2\Phi_i\Phi^{\dag i}+\left[
2\lambda_1\epsilon^{ijk}\Phi_i
d_{Rj}t_{Rk}+2\lambda_2\epsilon^{ijk}\Phi_i
\left(q_{Lj}Q_{Lk}\right)+{\rm h.c.}\right],
\eeq
\beq
\lag_{\rm eff}^{3}%
&=&-\frac{|\lambda_1|^2}{M^2}
\left[\mathcal{O}^8 _{V_dV_t}+\mathcal{O}^8 _{A_dA_t}
  +\mathcal{O}^8_{A_dV_t}+\mathcal{O}^8 _{V_dA_t}
  \right]\no\\
&-&\frac{|\lambda_2|^2}{M^2}
\left[\mathcal{O}^8 _{V_dV_t}+\mathcal{O}^8 _{A_dA_t}
  -\mathcal{O}^8_{A_dV_t}-\mathcal{O}^8 _{V_dA_t}\right]\no\\
    \no\\
&+&\frac{3\Re\left[\lambda_1\lambda^*_2\right]}{M^2}
\left[\mathcal{O}^8_{S_dS_t}+\mathcal{O}^8_{P_dP_t}\right]+\frac{3\Im\left[\lambda_1\lambda^*_2\right]}{M^2}\left[\mathcal{O}^8_{S_dP_t}+\mathcal{O}^8_{P_dS_t}\right]\no\\
&+&\frac{\Re\left[\lambda_1\lambda^*_2\right]}{4M^2}\left[\mathcal{O}^8_{T_dT_t}+\mathcal{O}^8_{T'_dT'_t}\right]+\frac{\Im\left[\lambda_1\lambda^*_2\right]}{4M^2}\left[\mathcal{O}^8_{T_dT'_t}+\mathcal{O}^8_{T'_dT'_t}\right]
\no\\
&+&\left(\mathcal{O}^8\to-\frac{1}{\sqrt{2}}\,\mathcal{O}^1\right).
\eeq
As in the case of the $q_LQ_L$ coupling of the $(\bar
6,1)_{-\frac{1}{3}}$ representation, we simplify the analysis by
imposing the conservative flavor physics constraint
$|\lambda_2|^4\left(M/{\rm TeV}\right)^{-2}<10^{-2}$.
\item {\bf Color-triplet weak-triplet $\Phi\sim(3,3)_{-{1}/{3}}$}
\beq
\lag^{(3,3)_{-{1}/{3}}}=-M^2\Phi_i\Phi^{\dag i}+
\left[2\lambda\epsilon^{ijk}\Phi_i\cdot\left(q_{Lj}Q_{Lk}\right)+{\rm h.c.}\right],
\eeq
\beq
\lag_{\rm eff}^{3}&=&-\frac{|\lambda|^2}{M^2}
\left[\mathcal{O}^8 _{V_uV_t}+\mathcal{O}^8 _{A_uA_t}-\mathcal{O}^8_{A_uV_t}
  -\mathcal{O}^8 _{V_uA_t} \right]
+\frac{1}{2}\times\left(u\leftrightarrow d\right)\no\\
&+&\left(\mathcal{O}^8\to-\frac{1}{\sqrt{2}}\,\mathcal{O}^1\right).
\eeq
\item {\bf Color-singlet weak-doublet $\Phi\sim(1,2)_{-{1}/{2}}$}
\beq\label{eq:lagsin}
\lag^{(1,2)_{-{1}/{2}}}=-M^2\Phi^\dag\Phi+
2\left[\lambda_1 q_L^{\dag i}\Phi t_{Ri}
  +\lambda_2 Q_L^{\dag i}\Phi u_{Ri}+\lambda_3 Q_L^{\dag i}\tilde\Phi
  d_{Ri}+{\rm h.c.}\right],
\eeq
\beq\label{eq:Leffsinglet}
\lag_{\rm eff}^{1}
&=&-\frac{|\lambda_1|^2}{M^2}
\left[\mathcal{O}^8 _{V_uV_t}-\mathcal{O}^8 _{A_uA_t}-\mathcal{O}^8_{A_uV_t}+\mathcal{O}^8 _{V_uA_t}\right]+\left(u\leftrightarrow d\right)\no\\
&-&\frac{|\lambda_2|^2}{M^2}
\left[\mathcal{O}^8 _{V_uV_t}-\mathcal{O}^8 _{A_uA_t}+\mathcal{O}^8_{A_uV_t}-\mathcal{O}^8 _{V_uA_t}\right]+\left(t\leftrightarrow b\right)\no\\
&-&\frac{|\lambda_3|^2}{M^2}
\left[\mathcal{O}^8 _{V_dV_t}-\mathcal{O}^8 _{A_dA_t}+\mathcal{O}^8_{A_dV_t}-\mathcal{O}^8 _{V_dA_t}\right]+\left(t\leftrightarrow b\right)\no\\
&+&\left(\mathcal{O}^8\to\frac{1}{2\sqrt{2}}\,\mathcal{O}^1\right).\eeq
Here, some comments are in order. As found in Section
\ref{sec:flavor}, flavor constraints imply that $\lambda_1\ll1$ and
$\lambda_2\lambda_3\ll1$, while $\lambda_2$ or $\lambda_3={\cal O}(1)$ in
order to produce the forward-backward asymmetry. We thus omit
in~\eqref{eq:Leffsinglet} terms proportional to $\lambda_1
\lambda_2^*$ generating same sign tops and single top production at
the LHC. When the spectrum of $\Phi$ breaks $SU(2)_L$, additional
terms of $\mathcal{O}(\lambda_1\lambda_2)$ adding to $t\bar t$
production also appear but are again suppressed and hence omitted. We
similarly neglect terms of ${\cal O}(\lambda_2 \lambda_3)$ in the
above. If one goes beyond the scope of the current work to incorporate
flavor symmetries, a sizable $\lambda_1$ may become allowed, and the
additional $SU(2)_L$ breaking terms should be considered as well.

\item {\bf Color-octet weak-doublet $\Phi\sim(8,2)_{-{1}/{2}}$}
\beq
\lag^{(8,2)_{-{1}/{2}}}&=&-M^2\Phi_a^\dag\Phi_a+
2\sqrt{6}\left(T^a\right)_i^j\left[\lambda_1 q_L^{\dag i}\Phi_at_{Rj}+
\lambda_2Q_L^{\dag i}\Phi_au_{Rj}
+\lambda_3Q_L^{\dag i}\tilde\Phi_ad_{Rj}+{\rm h.c.}\right],
\eeq
\beq
\lag_{\rm eff}^{8}
&=&\frac{|\lambda_1|^2}{M^2}
\left[\mathcal{O}^8 _{V_uV_t}-\mathcal{O}^8 _{A_uA_t}-\mathcal{O}^8_{A_uV_t}+\mathcal{O}^8 _{V_uA_t}
\right]+\left(u\leftrightarrow d\right)\no\\
&+&\frac{|\lambda_2|^2}{M^2}
\left[\mathcal{O}^8 _{V_uV_t}-\mathcal{O}^8 _{A_uA_t}+\mathcal{O}^8_{A_uV_t}-\mathcal{O}^8 _{V_uA_t} \right]+\left(t\leftrightarrow b\right)\no\\
&+&\frac{|\lambda_3|^2}{M^2}
\left[\mathcal{O}^8 _{V_dV_t}-\mathcal{O}^8 _{A_dA_t}+\mathcal{O}^8_{A_dV_t}-\mathcal{O}^8 _{V_dA_t} \right]+\left(t\leftrightarrow b\right)\no\\
&+&\left(\mathcal{O}^8\to-2\sqrt{2}\,\mathcal{O}^1\right).
\eeq
Similarly to the case of the color singlet isodoublet, $SU(2)_L$
breaking could in principle add terms to $\lag^{(8,2)_{-{1}/{2}}}$,
but flavor constraints imply that these terms are subleading.
\end{itemize}

We define the symmetric and antisymmetric partonic cross sections as
follows:
\beq
\hat\sigma_\pm=\int_{0}^1dc_\theta\frac{d\hat\sigma}{dc_\theta}
\pm\int_{-1}^0dc_\theta\frac{d\hat\sigma}{dc_\theta}.
\eeq
For heavy NP, the dominant contributions to the cross section arise
from interference with the SM:
\beq\label{eq:int}
\hat\sigma_{+}^{{\rm int}}=\frac{c_{VV}^{8}}{M^2}\,
\frac{2\alpha_s\,\beta_t}{9}\left(1-\frac{1}{3}\beta_t^2\right),\;\;\;
\hat\sigma_{-}^{{\rm int}}=\frac{c_{AA}^{8}}{M^2}\,
\frac{\alpha_s\,\beta_t^2}{9}.
\eeq
with $\beta_t^2=1-4m_t^2/\tilde s$. We take $m_t=172.5$ GeV.

Collecting the contributions from all of the operators, the
differential partonic $t\bar t$ production cross section is:
\beq\label{eq:cs}
\frac{d\hat\sigma}{dc_\theta}&=&
\frac{\alpha_s\beta_t\left(M^2+\frac{\tilde s}{2}-m_t^2
    +\eta\frac{\tilde s\beta_t}{2}c_\theta\right)}
{9\left[\left(M^2+\frac{\tilde s}{2}-m_t^2 +\eta\frac{\tilde
        s\beta_t}{2}c_\theta\right)^2+M^2\Gamma^2\right]}
\left[\left(1-\frac{\beta_t^2}{2}\right)c_{VV}^{8}+\beta_tc_{AA}^{8}\,c_\theta+\frac{\beta_t^2}{2}c_{VV}^{8}\,c_\theta^2\right]\\
&+&\frac{\tilde s\beta_t}{144\pi\left[\left(M^2+\frac{\tilde
        s}{2}-m_t^2+\eta\frac{\tilde
s\beta_t}{2}c_\theta\right)^2+M^2\Gamma^2\right]}\times\Sigma_{1,8}\no\\
&&\Big\{\beta_t^2c_V^2+2\left(1-\beta_t^2\right) \left(c_{VV}^2+c_{AV}^2\right)+c_{PP}^2+c_{SP}^2+\beta_t^2\left(c_{SS}^2+c_{PS}^2\right)+4\left(c_T^2+2c_{TT}c_{T'T'}+2c_{TT'}c_{T'T}\right)\left(1-\beta_t^2\right)\no\\
&+&2\left[2\left(c_{AV}c_{VA}+c_{VV}c_{AA}\right)-\left(c_{SS}+c_{PP}\right)\left(c_{TT}+c_{T'T'}\right)-\left(c_{SP}+c_{PS}\right)\left(c_{T'T}+c_{TT'}\right)\right]\beta_tc_\theta\no\\
&+&\left[c_V^2+8\left(c_T^2+2c_{TT}c_{T'T'}+2c_{TT'}c_{T'T}\right)\right]\beta_t^2c_\theta^2
\Big\},\no
\eeq
where
\beq
\eta=\Big\{\ba{c}+1,\;u{\rm -channel},\\-1,\;t{\rm -channel}.\ea
\eeq
We defined
\beq
c_V^2&=&c_{VV}^2+c_{AA}^2+c_{AV}^2+c_{VA}^2\no\\
c_T^2&=&c_{TT}^2+c_{T'T'}^2+c_{T'T}^2+c_{TT'}^2.\eeq
Lastly, in writing \[\Sigma_{1,8}\left\{\ldots\right\},\]
we intend that the term in curly brackets should be summed over the two orthogonal color contractions defined in Eq.~(\ref{eq:colbasis}).

\end{appendix}

\vspace{0.5cm}
\noindent{\Large \bf Acknowledgments}\\
We thank C\'edric Delaunay, Oram Gedalia, Amnon Harel, Alex Kagan,
Gilad Perez, Pedro Schwaller, Sean Tulin, Lidija Zivkovic and Jure Zupan for helpful
discussions. YN is the Amos de-Shalit chair of theoretical physics and
supported by the Israel Science Foundation (ISF)
under grant No.~377/07, by the German-Israeli foundation for
scientific research and development (GIF), and by the United
States-Israel Binational Science Foundation (BSF), Jerusalem,
Israel.
%


\end{document}